\let\emph\textit
\documentclass[sigconf]{acmart}
\usepackage{listings}
\AtBeginDocument{%
  }

\setcopyright{acmcopyright}
\copyrightyear{2024}
\acmYear{2024}
\setcopyright{acmlicensed}\acmConference[ASE '24]{39th IEEE/ACM International Conference on Automated Software Engineering }{October 27-November 1, 2024}{Sacramento, CA, USA}
\acmBooktitle{39th IEEE/ACM International Conference on Automated Software Engineering (ASE '24), October 27-November 1, 2024, Sacramento, CA, USA}
\acmDOI{10.1145/3691620.3694997}
\acmISBN{979-8-4007-1248-7/24/10}
\usepackage{multirow}
\usepackage{mdframed}
\usepackage{enumitem}
\usepackage{graphicx}
\acmConference[ASE 2024]{the 39th IEEE/ACM International Conference on Automated Software Engineering}{October 27--November 01,
  2024}{Sacramento, United States}

\usepackage{float}
\usepackage{threeparttable}
\usepackage{tcolorbox}

\newcommand{\find}[1]{
\begin{tcolorbox}[leftrule=1mm,toprule=0mm,bottomrule=0mm,left=1pt,right=2pt,top=2pt,bottom=2pt] %
#1
\end{tcolorbox}
}

\begin{document}

\title{Instructive Code Retriever: Learn from Large Language Model's Feedback for Code Intelligence Tasks}

\author{Jiawei Lu\footnotemark[1]}
\affiliation{%
  \institution{The State Key Laboratory of Blockchain and Data Security, Zhejiang University}
  \city{Hangzhou}
  \country{China}
}
\email{jiaweilu@zju.edu.cn}

\author{Haoye Wang\footnotemark[1]}
\affiliation{
  \institution{Hangzhou City University}
  \city{Hangzhou}
  \country{China}
}
\email{wanghaoye@hzcu.edu.cn}

\author{Zhongxin Liu\footnotemark[2]\footnotemark[3]}
\affiliation{
  \institution{The State Key Laboratory of Blockchain and Data Security, Zhejiang University}
  \city{Hangzhou}
  \country{China}
}
\email{liu\_zx@zju.edu.cn}

\author{Keyu Liang}
\affiliation{
  \institution{The State Key Laboratory of Blockchain and Data Security, Zhejiang University}
  \city{Hangzhou}
  \country{China}
}
\email{liangkeyu@zju.edu.cn}

\author{Lingfeng Bao}
\affiliation{
  \institution{The State Key Laboratory of Blockchain and Data Security, Zhejiang University}
  \city{Hangzhou}
  \country{China}
}
\email{lingfengbao@zju.edu.cn}

\author{Xiaohu Yang}
\affiliation{
  \institution{The State Key Laboratory of Blockchain and Data Security, Zhejiang University}
  \city{Hangzhou}
  \country{China}
}
\email{yangxh@zju.edu.cn}

\renewcommand{\shortauthors}{Lu et al.}
\newcommand{\hy}[1]{\textcolor{cyan} {hy : {#1}} }
\newcommand{\todo}[1]{\textcolor{red} {hy : {#1}} } 
\newcommand{\jw}[1]{\textcolor{violet} {jw : {#1}} } 
\begin{abstract}
Recent studies proposed to leverage large language models (LLMs) with In-Context Learning (ICL) to handle code intelligence tasks without fine-tuning.
ICL employs task instructions and a set of examples as demonstrations to guide the model in generating accurate answers without updating its parameters. While ICL has proven effective for code intelligence tasks, its performance heavily relies on the selected examples. Previous work has achieved some success in using BM25 to retrieve examples for code intelligence tasks. However, existing approaches lack the ability to understand the semantic and structural information of queries, resulting in less helpful demonstrations. Moreover, they do not adapt well to the complex and dynamic nature of user queries in diverse domains.
In this paper, we introduce a novel approach named Instructive Code Retriever (ICR), which is designed to retrieve examples that enhance model inference across various code intelligence tasks and datasets. We enable ICR to learn the semantic and structural information of the corpus by a tree-based loss function.
To better understand the correlation between queries and examples, we incorporate the feedback from LLMs to guide the training of the retriever.
Experimental results demonstrate that our retriever significantly outperforms state-of-the-art approaches.
We evaluate our model's effectiveness on various tasks, i.e., code summarization, program synthesis, and bug fixing. Compared to previous state-of-the-art algorithms, our method achieved improvements of 50.0\% and 90.0\% in terms of BLEU-4 for two code summarization datasets, 74.6\% CodeBLEU on program synthesis dataset, and increases of 3.6 and 3.2 BLEU-4 on two bug fixing datasets. 

\end{abstract}


\maketitle

\renewcommand{\thefootnote}{\fnsymbol{footnote}}
\footnotetext[1]{Equal Contribution.}
\footnotetext[2]{Corresponding Author.}
\footnotetext[3]{Also with Hangzhou High-Tech Zone (Binjiang) Institute of Blockchain and Data Security}

\section{Introduction}

Code intelligence tasks aim to automate the analysis, understanding, and generation of code. Representative code intelligence tasks include code summarization, bug fixing, and program synthesis.
These tasks are highly valued because of their potential to mitigate the time-consuming nature of manual efforts, such as annotation composition, debugging, and code creation \cite{tufano2021towards,li2022automating,lu2023improving}.
As the complexity of software systems continues to increase, there is a growing need for efficient and automated approaches to understanding, modifying, and improving source code \cite{rigby2013convergent,leclair2020improved,tufano2019empirical,gulwani2017program}. 

In recent years, pre-trained models like CodeBERT \cite{feng2020codebert}, GraphCodeBERT \cite{guo2020graphcodebert} and CodeT5 \cite{wang2021codet5} have made significant strides across various code intelligence tasks. They are pre-trained on large-scale code corpora with diverse objects.
However, when dealing with various code intelligence tasks, these pre-trained models need to be fine-tuned separately on datasets for each task, limiting their generalizability. Additionally, the parameter sizes of these pre-trained models are relatively, restricting the knowledge they can capture \cite{wei2022emergent} and thereby preventing them from achieving satisfactory performance across multiple tasks simultaneously \cite{li2023unified}.

\begin{figure}[t]
  \centering
  \includegraphics[width=\linewidth]{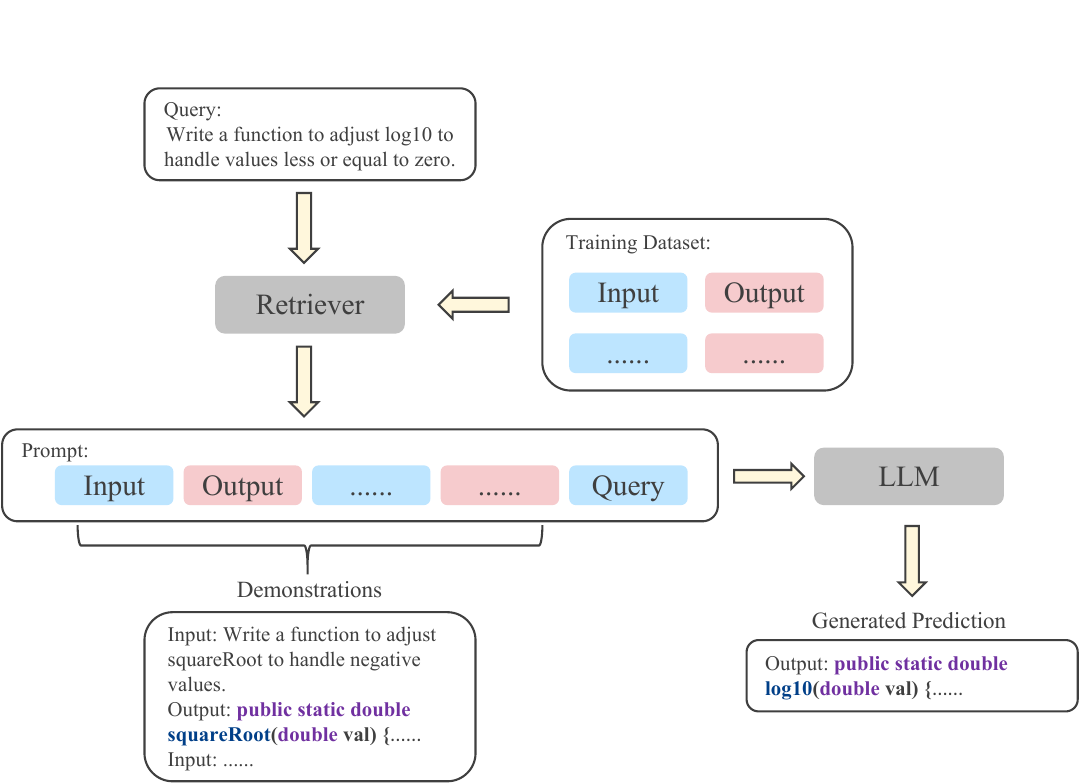}
  \vspace{-0.2cm}
  \caption{An Example of In-Context Learning on Program Synthesis Task.}
  \label{fig:icl}
\vspace{-0.2cm}
\end{figure}

Recently, large language models (LLMs) have attracted a lot of attention due to their impressive effectiveness on diverse tasks \cite{huang2023empirical,zhout2023devil,xia2023plastic}. In contrast to traditional pre-trained models, LLMs increase parameter counts tenfold or even hundredfold \cite{brown2020language} and have been trained on massive datasets of both natural language and programming languages \cite{roziere2023code}. Given appropriate prompts, LLMs demonstrate exceptional proficiency across diverse code intelligence tasks, obviating the necessity for parameter fine-tuning.
However, prior research indicates that LLMs exhibit sensitivity to input prompts \cite{qiao2022reasoning}. 
To better guide LLMs in performing code intelligence tasks, several studies have explored In-Context Learning (ICL)~\cite{gao2023makes,ahmed2024automatic,nashid2023retrieval}, which provides LLMs with several input-output pairs to showcase how to follow the instruction. As illustrated in Figure \ref{fig:icl}, the retriever retrieves a series of demonstrations from the training set and adds them before the query, thereby prompting the LLM to output the correct answer.
Instead of fine-tuning the model for a task, ICL can achieve efficient knowledge transfer with just a few examples, making it effective for handling diverse tasks and adapting to rapidly changing environments.

Previous researches show that the selection of demonstrations is crucial to the performance of In-Context Learning (ICL)~\cite{liu2021makes,gao2023makes}.
Gao et al. \cite{gao2023makes} find that BM25 \cite{robertson1994some} can effectively retrieve ICL examples for various code intelligence tasks.
Ahmed et al.~\cite{ahmed2024automatic} improve their prompts with designed semantic facts for code summarization after BM25 retrieved examples. They name their approach as ASAP.

However, BM25-based retrievers have limited capabilities for understanding rich structural information in programming languages, which are useful for code intelligence tasks \cite{zhu2021syntax}. Additionally, they lack the essential domain knowledge and semantic understanding ability to comprehend the correlation between queries and examples. These capabilities are important for code intelligence tasks because programs in different forms may have similar functionality. Moreover, ASAP is only applicable to specific datasets and tasks.

In this research, we present a novel approach named \textbf{I}nstructive \textbf{C}ode \textbf{R}etriever (ICR) to address the aforementioned limitations.
We use a combined encoding of text and syntax trees to calculate the similarity of examples. Then we propose a tree-based loss function to train the retriever, allowing the retriever to consider both semantic and structural information.
Additionally, we utilize feedback from the LLM to rank the quality of examples, enabling the retriever to leverage the LLM's domain knowledge and semantic understanding ability through contrastive learning.

Specifically, in the training phase, ICR initially utilizes syntax trees to capture structural information from the queries and examples.
Then we leverage the inherent capabilities of LLMs to discern the usefulness of examples for each query, thereby obtaining both positive and negative instances for training a retriever capable of selecting superior examples. 
Finally, we train the retriever iteratively by minimizing a customized tree-based loss between the retriever's similarity and the sequence feedback from the large model.
In the testing phase, we utilize the trained retriever to retrieve valuable examples. These instances are then used to construct prompts, which are employed to enhance the performance of LLMs with respect to the given code intelligence tasks.

We evaluate our proposed retriever ICR on three code intelligence tasks, i.e., code summarization, program synthesis, and bug fixing, using eight datasets in total.
Compared to previous state-of-the-art algorithms, ICR achieves improvements of 50.0\% and 90.0\% on the Java and Python datasets in terms of BLEU-4 for code summarization, 74.6\% in terms of CodeBLEU on program synthesis, and increases of 3.6 BLEU-4 and 3.2 BLEU-4 on two datasets for bug fixing with the base model GPT-Neo-2.7B. 
When transferring ICR trained on GPT-Neo-2.7B to Code Llama-13B, ICR improves over the state-of-the-art baselines by up to 18.3\%, 20.5\%, and 1.6\% on the code summarization, program synthesis, and bug fixing tasks on BLEU-4, respectively.
Additionally, ICR outperformed state-of-the-art methods on code summarization datasets in three languages that ICR had not been trained on. Moreover, we found that when we use high-quality examples, the order of examples has little impact on performance.

To summarize, the contributions of our work are:

\begin{itemize}
    \item We propose a novel retriever to retrieve valuable demonstrations that can help large language models accomplish code intelligence tasks. To the best of our knowledge, this is the first method that utilizes large language model feedback for contrastive learning to enhance the performance of ICL in code intelligence tasks.
    \item To utilize the structural information from the corpus, we introduce a tree-based model loss function tailored specifically for code intelligence tasks.
    \item We evaluated the effectiveness of our approach across multiple programming languages and various code intelligence tasks. Comprehensive experimental results demonstrate the effectiveness of our proposed approach.
\end{itemize}

\section{Motivation}

\begin{figure}[H]
  \centering
  \includegraphics[width=\linewidth]{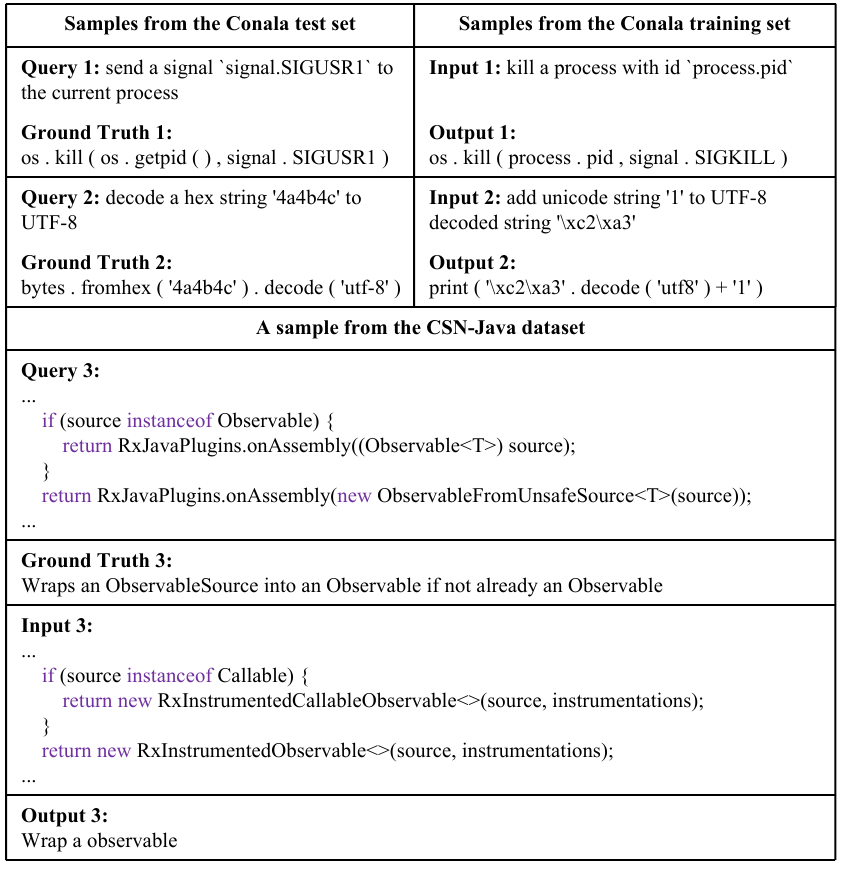}
  \vspace{-0.3cm}
  \caption{Examples that BM25 will fail in such cases.}
  \label{fig:bm25_fail}
\end{figure}

For code intelligence tasks, existing ICL methods~\cite{gao2023makes, ahmed2024automatic} use BM25 \cite{robertson1994some} as the retrieval engine to retrieve examples. 
The upper left part of Figure~\ref{fig:bm25_fail} presents two test samples collected from the Conala dataset of the program synthesis task.
The upper right part shows a training sample for each test sample.
For the first case, the queries of the test sample and the training sample, i.e., Query 1 and Input 1, differ significantly in the used tokens.
Therefore, BM25 will not retrieve this training sample for Query 1.
However, the structure and functionality of their outputs, i.e., Ground Truth 1 and Output 1, are very similar, both sending a signal to a specified process.
So this training sample can be a valuable demonstration for LLMs.
For the second case, Query 2 and Input 2 are lexically similar, so BM25 tends to retrieve this training sample as a demonstration.
However, they are semantically different, resulting in Ground Truth 2 and Output 2 differ significantly. 
Using such a demonstration for ICL may mislead LLMs.
These two cases show that domain knowledge and semantic understanding ability are important when retrieving demonstrations.

\begin{figure*}[th]
  \centering
  \includegraphics[width=0.9\linewidth]{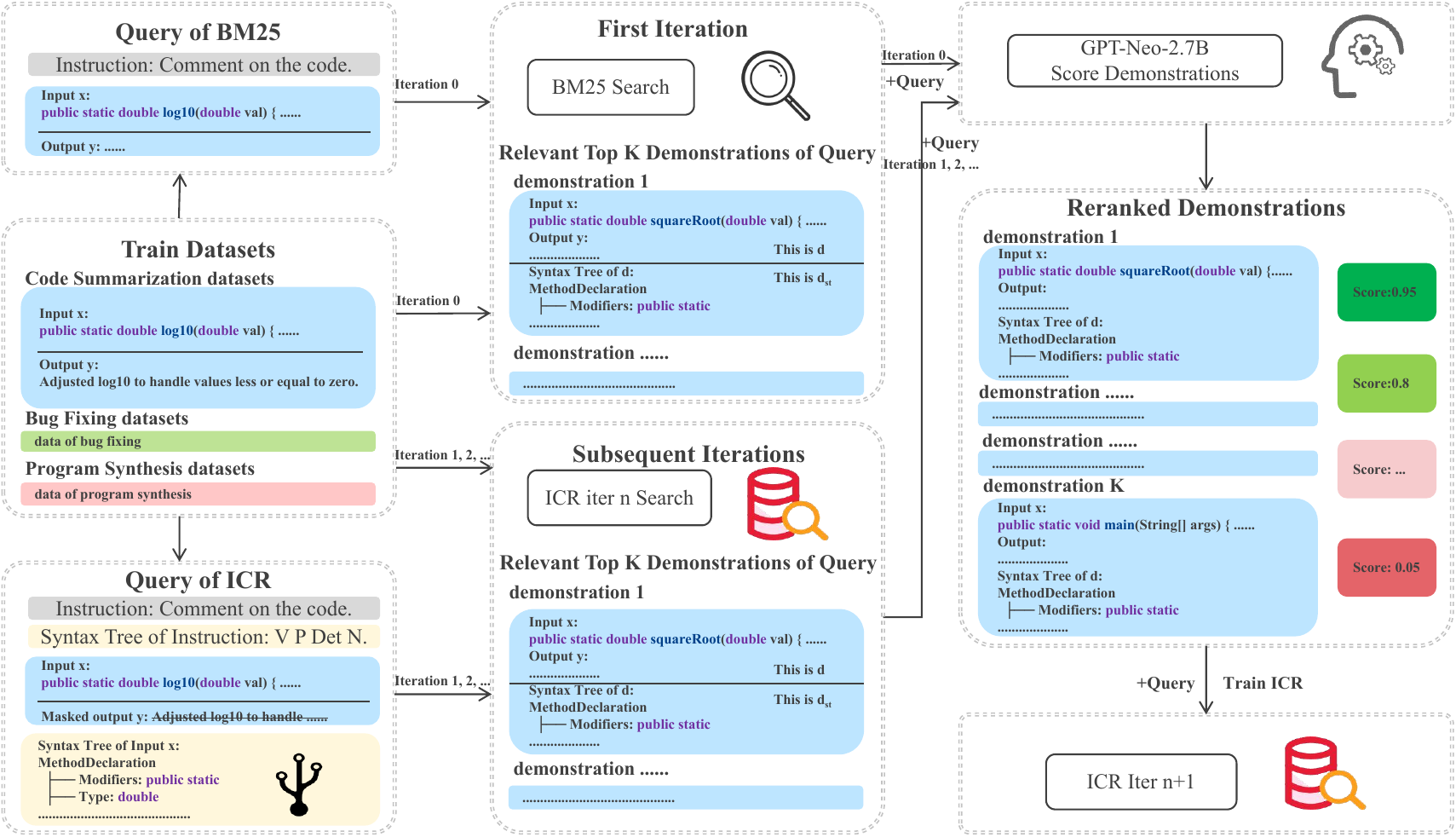}
  \caption{The Training Workflow of \textsc{ICR}.}
  \label{fig:overview}
\end{figure*}

The lower part of Figure~\ref{fig:bm25_fail} shows a test sample and a training sample collected from the CSN-Java dataset of the code summarization task.
Due to space limitation, we omit some code snippets in Query 3 and Input 3.
We can see that the input code snippets of the two samples, i.e., Query 3 and Input 3, share similar code structures and their code summaries, i.e., Ground Truth 3 and Output 3, are also alike. 
This training sample may help LLM generate a good summary for Query 3.
However, BM25 overlooks this example since it does not take structural information into account.

To address these issues, we propose the Instructive Code Retriever (ICR). 
We use contrastive learning to train the ICR based on the feedback from an LLM, leveraging the semantic understanding ability of the LLM to better analyze the input.
To incorporate the structural information, we introduce a tree-based loss computation approach which is applicable to code tasks.
Details of ICR will be described in the following section.

\section{Instructive Code Retriever}
\subsection{Overview}
In this section, we will outline the training and inference process of ICR. The retriever aims to retrieve examples from the training set that help the large language model in the inference process. 

As illustrated in Figure \ref{fig:overview}, the origin input is some sample from the training set.
In preprocessing and preparation, we first concat task-specific instructions to the origin input. Subsequently, we construct syntax trees for both program and natural language. After preprocessing, we get the input form shown in the lower left part of Figure \ref{fig:overview} as a query.
For each query, the first step is to retrieve a set of relevant examples from the corpus.
After that, we use a large language model to score whether examples would be helpful for LLM inference. 
Based on the feedback from the language model, we use contrastive learning to learn patterns of the examples that are more useful for the given task. We continuously optimize the training data and iteratively train the retriever.
After training, ICR can identify which example is helpful for LLM inference and which is not for a given sample. 
Once the retriever is trained, during inference, for each test set sample, we retrieve examples most beneficial for the large language model inference process. Finally, we use this sample and retrieved examples as input to the LLM for inference.

\subsection{Preprocessing and Preparation}
In this section, we will describe the preparatory process required for training ICR.

\textbf{Building Instructions of Tasks.}
ICR is a unified framework for multiple tasks. To enable the model to distinguish between different tasks, we need to incorporate task-specific instructions into the query. Task instructions are short texts describing the task objectives. For example, in the code summarization task, a task instruction could be "Comment on the code." We concatenate the task instruction with the original input for model training, forming the ICR model query in the following format: $(x, y)=T_i \oplus  (m, n)$, where $T_i$ is the task instruction, $(m, n)$ is the input-output data pair from training datasets, and $\oplus$ is the concatenation operator.

\textbf{Building Syntax Trees.}
Traditional bi-encoder architectures typically employ encoders $E_q$ and $E_d$ to encode input and example separately \cite{karpukhin2020dense}, followed by the computation of their similarity scores. 
The objective of training ICR is to distinguish which examples are highly relevant to the query. 
The most intuitive approach is to encode both the query and the examples directly and then compute the similarity.
However, programming languages inherently possess rich structural information, which is as important as textual information for distinguishing code \cite{zhu2021syntax}. 
Considering that natural language $T_i$ can also be represented as a syntactic tree, the contribution of structural information to the retriever's effectiveness may be augmented~\cite{an2023context}. 
Therefore, we introduce the syntactic tree structure similarity to the traditional bi-encoder architecture to assist the retriever in better understanding the similarity between queries and examples.

The query and examples include both programming languages and natural languages e.g., the input $m$ of program synthesis tasks is a natural language description, and instructions are also in natural language. 
In our approach, we adopt the `javalang' \cite{javalang} and `ast' \cite{python-ast} libraries to derive the abstract syntax tree of code, while for the natural language query, we rely on the `spacy' \cite{spacy} library to extract its syntax tree.
Finally, we get the preorder traversal of the syntax tree of query $x_\text {st}$ and the syntax tree of example $d_\text {st}$. 
The extracted syntax trees will be used in several places, e.g. the model training, the similarity calculation between the query and the example, and the calculation of the loss function. We will describe the details of these processes in the following sections.

\textbf{Selecting Relevant Examples for Each Sample.}\label{sec:retrieve_examples}
The primary objective of ICL is to procure high-quality examples for reference by LLM. To achieve this, the retriever plays a pivotal role in selecting exemplary instances. Since the relevant examples found in the previous section may not aid in model inference, we need to use contrastive learning to enable the retriever to identify helpful examples.
We need to score examples of each sample for contrastive learning. However, the computational expense associated with scoring the entire training set scales quadratically with its size, rendering such a task both impractical and redundant. 
Hence, our preliminary step is to roughly screen to find some examples that are similar to the given sample.
For each query from the training set, we use only the top K examples (except the query itself) initially retrieved as training data, enabling the model to learn patterns of examples that are useful for LLM inference.
This reduces unnecessary computational burden while ensuring the acquisition of high-quality instances for the training process.

\begin{figure}[h]

  \centering
  \includegraphics[width=0.85\linewidth]{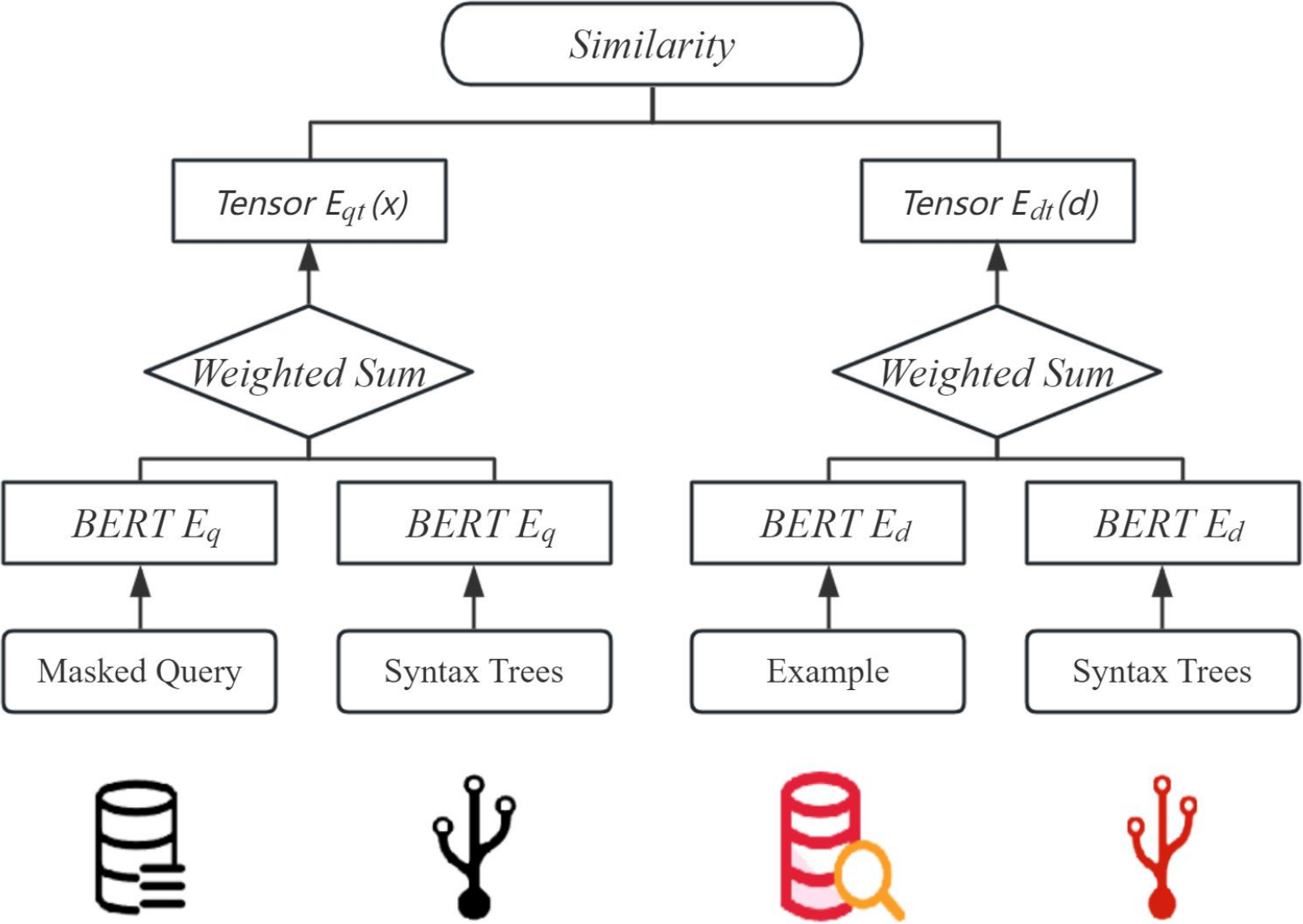}
  \vspace{-0.2cm}
  \caption{The Architecture of ICR.}
  \label{fig:architecture}
\vspace{-0.4cm}
\end{figure}

Specifically, we divide our process of selecting relevant examples for each sample into the following two scenarios:
\begin{itemize}
    \item \textbf{BM25: }In iter 0, when our retriever is not yet trained, we search for the top K examples (i.e., input-output pair $d$) with the highest BM25 scores for each sample in the training set. Noting that we are currently processing the training set and can fully utilize ground truth knowledge, to narrow down the scope and select the most relevant examples for a sample, we calculate the similarity between the query of iter 0 (i.e., $(x,y)$) and other examples in the training set.
    \item \textbf{ICR: }In iter 1, 2, ..., n, once our retriever is trained, we iteratively use the trained retriever to find the top K examples (i.e., input-output pair $d$) with the highest scores for each masked query $(x,-) = T_i \oplus  (m, -)$ in the training set. At this stage, the ground truth $n$ of the query is masked to ensure consistency between training and testing phases. The inputs for ICR include the query itself and its syntax tree, as well as the examples and their syntax trees.  Figure \ref{fig:architecture} shows the model architecture, and the training process will be introduced in \ref{training}.
\end{itemize}

\textbf{Scoring for Relevant Examples.} 
We aim to train ICR to distinguish how useful each relevant example is for handling the given query.
To achieve this goal, we need a reference ranking of the relevant examples mentioned above to guide the training.
We propose to leverage LLMs to obtain such reference ranking.
Specifically, for a given sample $(x, y)$ and its set of examples $(d_1, d_2, ..., d_k)$, we define the score of an example as the probability that the LLM $G$ outputs the ground truth result $y$ based on the input $x$ and the example $d_i$, as follows:
\begin{equation}
S\left(d_{i}\right) = \operatorname{P}_{G}\left(y \mid d_{i}, x\right)
\end{equation}
Based on the scores of examples, we obtain the ranking $r(d_i)$ of examples, which will guide the training of the retriever in subsequent sections. The function for $r(d_i)$ is as follows:

\begin{equation}
r\left(d_{i}\right)=\operatorname{rank}\left(j<i \mid S\left(d_{j}\right)>S\left(d_{i}\right)\right)
\end{equation}
Examples with higher scores are ranked higher and have smaller $r(d_i)$.

\subsection{Training ICR}\label{training}

ICR is responsible to calculate the similarity or relevance between a masked query and an example. 
Figure~\ref{fig:architecture} illustrates the architecture of ICR.
Given a masked query $x$ an example $d$, ICR first encodes $x$ and its syntax tree $x_{st}$ using a BERT encoder $E_q$, and encodes $d$ and its syntax tree $d_{st}$ using another BERT encoder $E_d$. 
Note that $d$ contains both the input and the output of the example, and each BERT encoder outputs the contextual embedding of the `CLS` token as the embedding of the inputs.
Then, we aggregate the embeddings of $x$ and $x_{st}$ and the embeddings of $d$ and $d_{st}$, respectively, as follows:

\begin{equation}
E_{q t}(x) = \alpha_{1} E_{q}(x)+\beta_{1} E_{q}\left(x_{\text {st}}\right)
\end{equation}
\begin{equation}
E_{d t}(d) = \alpha_{2} E_{d}(d)+\beta_{2} E_{d}\left(d_{\text {st}}\right)
\end{equation}
where $\alpha_{1}$, $\beta_{1}$, $\alpha_{2}$, $\beta_{2}$ are four trainable parameters.
Finally, we regard the dot product between the vectors $E_{q t}(x)$ and $E_{d t}(d)$ as the similarity score between the query and the example, as follows:

\begin{equation}
\operatorname{sim_{tree}}(x, d)=E_{q t}(x)^{\top}E_{d t}(d)
\end{equation}

The loss function for training consists of two parts:

\textbf{In-Batch-Tree Loss: }To enable the retriever to distinguish the best examples from others, we get inspiration from the loss function of Karpukhin et al. \cite{karpukhin2020dense}. We modify the in-batch loss function by adding structural information to ensure it aligns with the requirements of code-related tasks. We take into account structural similarity by using the similarity function defined above. We name the loss function as the in-batch-tree loss. The definition of the in-batch-tree loss function is as follows:

\begin{equation}
\mathcal{L}_{b t}=-\log \frac{e^{\operatorname{sim_{tree}}\left(x, d^{+}\right)}}{e^{\operatorname{sim_{tree}}\left(x, d^{+}\right)}+\sum_{d^{-} \in \mathbb{Z}} e^{\operatorname{sim_{tree}}(x, d^{-})}}
\end{equation}
Where the positive example $d^{+}$ denotes the highest-scoring example of $x$ within a batch, while $\mathbb{Z}$ encompasses all other examples of $x$ and examples of other queries within this batch. Notice that the negative sample contains not only suboptimal examples of the current query but also examples of other queries in the same batch (irrelevant examples).
The in-batch-tree loss function adopts the principle of contrastive learning, empowering the retriever to pinpoint the example within a batch that offers maximal assistance in inferring the ground truth.

\begin{figure*}[h]
  \centering
  \includegraphics[width=0.75\linewidth]{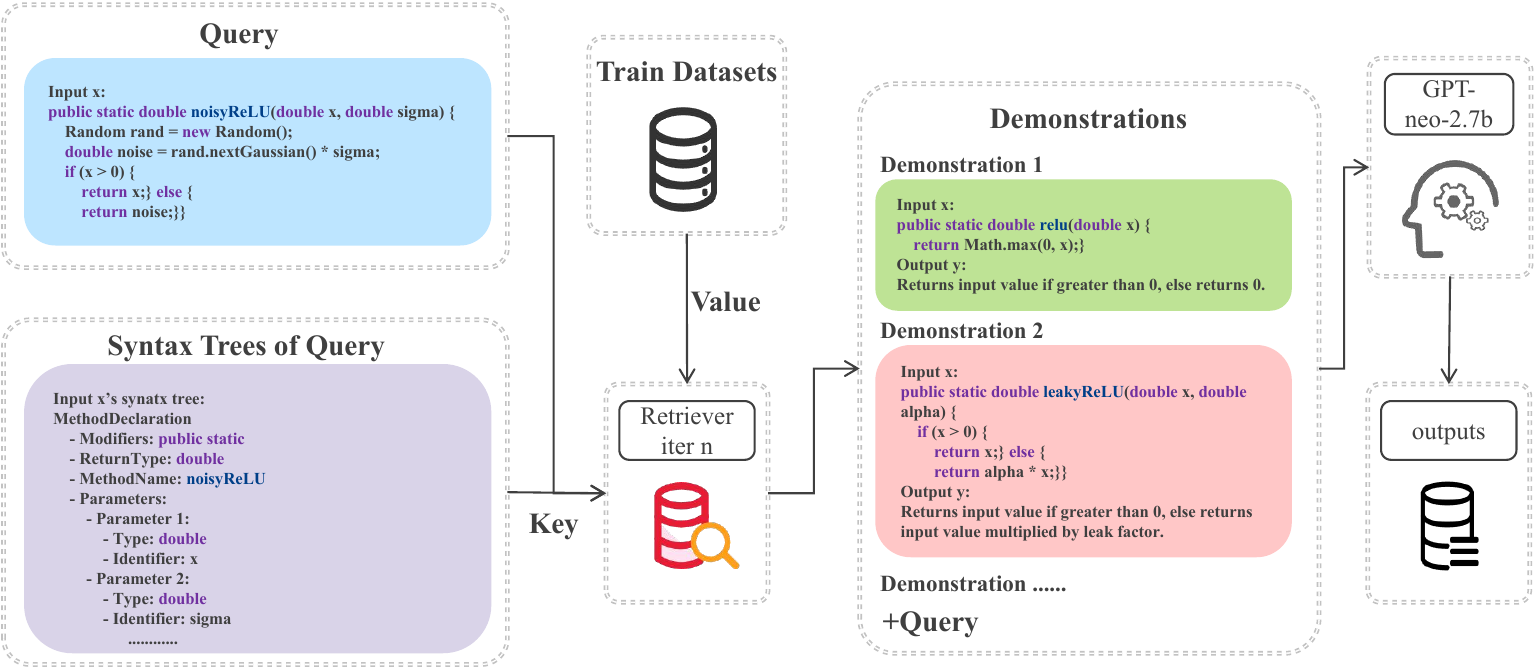}
  \vspace{-0.2cm}
  \caption{The Inference Workflow of \textsc{ICR}.}
  \label{fig:test}
\vspace{-0.4cm}
\end{figure*}

\textbf{Ranking-Tree Loss: } The in-batch-tree loss facilitates the retriever's discernment between the highest-quality examples and their suboptimal counterparts. However, the ordering among sub-optimal examples also warrants consideration. Therefore, drawing inspiration from Burges et al. \cite{burges2010ranknet} and Li et al. \cite{li2023unified}, we introduce a ranking-tree loss function, defined as follows:

\begin{equation}
\mathcal{L}_{r t}=\sum_{d_{i}, d_{j} \in D} r * \log \left(1+e^{\operatorname{sim_{tree}}\left(x, d_{j}\right)-\operatorname{sim_{tree}}\left(x, d_{i}\right)}\right)
\end{equation}
\begin{equation}
r=\max \left(0, \frac{1}{r\left(d_{i}\right)}-\frac{1}{r\left(d_{j}\right)}\right)
\end{equation}
This loss function optimizes $sim(x, d_i)>sim(x, d_j)$ when the ranking of $d_i$ is higher than that of $d_j$, i.e., $r(d_i)<r(d_j)$. Consequently, the retriever calculates higher similarity scores for examples ranked higher. The greater the disparity in rankings $r$, the larger the optimization, thus enlarging the gap in similarity between $d_i$ and $d_j$.

The overall tree-based loss function is as follows:

\begin{equation}
\mathcal{L}=\gamma_1 * \mathcal{L}_{b t}+\gamma_2 * \mathcal{L}_{r t}
\end{equation}

where $\gamma_1$ and $\gamma_2$ are hyper-parameters.

In summary, our training objective is to enable the retriever to identify which examples provide the most assistance in obtaining ground truth from the input in the training set. We achieve this by obtaining the ranking of example quality through feedback from the large model on the training set examples. We train the retriever based on this ranking.

\subsection{Inferencing for Specific Tasks}\label{inference}

During inference, we pre-encode all training set examples using $E_{dt}$. This is because our retriever relies on dot product similarity to rank examples.
Precomputing $E_{dt}$ helps avoid redundant computations. For each test example, we get its masked query $(x_{test}, -)$ and compute its encoding $E_{qt}$, and then use the FAISS \cite{johnson2019billion} library to search for a series of examples in the training data that maximize the inner product $sim_{tree}(x_{test}, d)$. We then sort the examples from highest to lowest, resulting in a sequence $D=(d_1, d_2 ... d_L)$. The final input prompt $P$ provided to the model depends on the maximum input length $|C_i|$ and the maximum output length $|C_o|$ of the model, ensuring that $|P|={\textstyle \sum_{i=1}^{L}|d_i|}+...+|x_{test}|+|C_o|<|C_m|$. We set a hyperparameter $|C_m|$ and ensure $|C_m|<|C_i|$, instead of using a fixed number of examples. This effectively avoids the example truncation issue. and such that $P$ is the maximal length among all $P_i$ satisfying the condition. We place examples with higher similarity closer to the input. The input prompt format for LLM is as follows:

\begin{equation}
Instruction\parallel Exemplar_1 \parallel \dots \parallel Input\Rightarrow Prompt
\end{equation}

With this prompt, we can input it into the large model and obtain the final output. 

\section{Experimental Design}

In this section, we describe the experimental setup that we follow to evaluate the effectiveness of our ICR. Our experiments aim to answer the following research questions:

\textbf{RQ1: How effective is ICR enhance the LLM compared to existing state-of-the-art methods?}
The primary goal of In-context Learning is to enhance the performance of language models in solving specific tasks. Therefore, we selected two language models of different sizes and conducted comparative experiments to investigate the effectiveness of ICR. In this section,  we compared ICR with state-of-the-art In-context Learning approaches for code intelligence tasks: BM25~\cite{gao2023makes} and \textit{ASAP}~\cite{ahmed2024automatic}.

\textbf{RQ2: What are the impacts of different modules in the training of ICR?}
In RQ2, we will individually dissect the impact of two components, i.e. structural information and iterative training based on LLM feedback, on the performance of our model, serving as ablation studies.

\textbf{RQ3: Can our ICR achieve satisfactory performance on untrained code datasets compared with BM25 and ASAP \cite{ahmed2024automatic}?}
In RQ3, we will compare the performance of JavaScript, PHP and Ruby datasets, which are not used during ICR training, to demonstrate the generalization ability of our model on untrained datasets.

\vspace{-0.2cm}
\begin{table}[h]
\centering
\caption{Statistics of the benchmark datasets.}
\label{tab: datasets}
\vspace{-0.2cm}
\resizebox{\columnwidth}{!}{
\begin{tabular}{l|llll}
\hline
                                                     & Datasets    & Train  & Validation & Test  \\ \hline
                                                     & CSN-Java    & 164923 & 5183  & 10955 \\
\multirow{3}{*}{Code Summarization}                 & CSN-Python  & 251820 & 13914 & 14918 \\ 
                                                     & CSN-JavaScript  & 58025 & 3885 & 3291 \\ 
                                                     & CSN-PHP  & 241241 & 12982 & 14014 \\ 
                                                     & CSN-Ruby  & 24927 & 1400 & 1261 \\ 
\hline                          & $B2F_{small}$  & 46680  & 5835  & 5835  \\
\multirow{-2}{*}{Bug Fixing} & $B2F_{medium}$ & 52364  & 6546  & 6545  \\ \hline
Program Synthesis                                    & Conala      & 2379   & -     & 500   \\ \hline
\end{tabular}}
\vspace{-0.2cm}
\end{table}

\subsection{Datasets}

Our experiment involves three code intelligence tasks.
Table \ref{tab: datasets} presents the statistics of the datasets we used in our experiments.

\textbf{Bug Fixing:}
The bug-fixing dataset used in our study is the popular B2F dataset collected by Tufano et al. \cite{tufano2019empirical} and included in the \textsc{CodeXGLUE} benchmark \cite{lu2021codexglue}. Following the original paper's setup, the B2F dataset is divided based on the length of code tokens into two parts: $\text{B2F}_{medium}$ and $\text{B2F}_{small}$. This dataset was also used by Gao et al. to evaluate the enhancement of LLM through ICL. Hence, we employ the B2F dataset for comparison with our proposed approach. The dataset consists of pairs of buggy code snippets and their corresponding fixed versions in Java, where each pair represents a bug-fixing commit extracted from GitHub repositories. We directly adopt the original dataset splits for our experiments.

\textbf{Program Synthesis:}
For the program synthesis task, we utilize the CoNaLa dataset~\cite{yin2018learning}. The CoNaLa dataset, short for Code/Natural Language (CoNaLa), is a widely used benchmark dataset designed for the task of program synthesis from natural language descriptions \cite{mishra2022lila,gao2023makes,zhuo2024ice}. This dataset consists of Python data collected from Stack Overflow. With its diverse range of programming scenarios, varying levels of complexity, and high-quality annotations, the CoNaLa dataset serves as a worth considering dataset. We directly adopt the original dataset splits for our experiments.

\textbf{Code Summarization:}
For the code summarization task, we employ the CodeSearchNet dataset \cite{husain2019codesearchnet}. This dataset contains millions of functions from open-source projects across multiple programming languages, making it popular for training and evaluating models on code summarization tasks. 
Due to its high quality, the CodeSearchNet dataset has also been included in \textsc{CodeXGLUE}.
Code summarization involves generating concise and accurate summaries of code functionality, comments, or documentation.  
This dataset provides a diverse range of code examples, enabling the development of models that can effectively summarize code across different programming languages and domains. We directly adopt the original dataset splits for our experiments.

\subsection{Metrics}
Since we evaluate the performance of ICR on multiple tasks that have different outputs, we used different metrics as follows.

\textbf{BLEU:}
BLEU \cite{papineni2002bleu} stands as a widely used metric in assessing bug fixing and code summarization tasks. It quantifies the resemblance between automatically generated text and reference text. BLEU evaluates this resemblance by analyzing the overlap of n-grams between the candidate text and multiple reference texts. Precision scores for n-grams in the generated text are computed, weighted by their frequency.

\textbf{ROUGE-L:}
ROUGE-L \cite{lin2004rouge} is a metric used to evaluate the quality of summaries by measuring the longest common subsequence (LCS) between the generated summary and a reference summary. ROUGE-L captures the sentence-level structure similarity by considering the sequence of words. This makes it particularly effective for assessing how well a summary retains the order and context of the information from the source text, providing a more nuanced understanding of its coherence and relevance. 

\textbf{chrF}
chrF \cite{popovic2015chrf} is a character-level evaluation metric, which measures the quality of outputs by comparing character n-grams between the system output and reference sentences. By operating at the character level, chrF effectively captures subtle differences in morphology, spelling, and word inflections, making it particularly valuable for evaluating translations in morphologically rich languages or low-resource scenarios. This metric calculates an F-score based on the precision and recall of character n-gram overlaps. 

\textbf{METEOR}
METEOR \cite{roy2021reassessing} is an evaluation metric designed to improve traditional metrics like BLEU by incorporating linguistic features such as stemming and synonymy. It aligns words between the system output and reference sentences, allowing for partial matches and better handling of varied expressions of meaning. METEOR balances precision and recall, with a slight preference for recall, to reward outputs that capture more reference content. Additionally, it imposes penalties for differences in word order, making it more sensitive to fluency and grammaticality. 

\textbf{CodeBLEU:}
CodeBLEU \cite{ren2020codebleu}, an adaptation of the BLEU metric, plays a crucial role in evaluating code generation models. By integrating abstract syntax trees, CodeBLEU enhances its evaluation capabilities, infusing code syntax into the assessment process. Furthermore, it leverages data flow to incorporate semantic information, thereby enriching the evaluation process. This integration enables a more comprehensive assessment of code generation models, considering both syntactic correctness and semantic fidelity in the generated code.

\textbf{CrystalBLEU}
CrystalBLEU \cite{eghbali2022crystalbleu} is an enhanced version of the traditional BLEU metric designed to improve machine translation evaluation by incorporating linguistic insights. While BLEU focuses on n-gram precision and exact word matches, CrystalBLEU extends this by considering syntactic structures and semantic equivalence, allowing it to account for variations like paraphrases and synonymy. This makes CrystalBLEU more effective in capturing the nuances of translation quality, especially in scenarios where different but equally valid translations may use varied word choices or syntactic structures.

\subsection{Base Models}

In this paper, we employ the ICL method to retrieve examples that enhance the performance of LLMs. 
In order to show our ICR can work with different LLMs, we selected two popular open-source LLMs. 
Since the Codex API (Code-Davinci-002), which was widely used in previous research \cite{gao2023makes,ahmed2024automatic}, has been deprecated by OpenAI and is no longer available, and other API models may undergo performance changes with version updates or be deprecated, we prefer to use open-source models to ensure the reproducibility of our work.
We respectively enhance each LLM with ICR and other baseline methods. 
The introductions to these two LLMs are as follows:

\textbf{GPT-Neo-2.7B}
GPT-Neo-2.7B \cite{black2021gpt} is an open-source language model developed by EleutherAI as a distilled version of OpenAI's GPT-3. With 2.7 billion parameters, GPT-Neo-2.7B offers substantial performance comparable to much larger models while being more computationally efficient. It is designed to perform a wide range of tasks, demonstrating strong generalization abilities across various domains without fine-tuning. This makes it particularly suitable for tasks requiring significant computational resources and those that benefit from a distilled model's efficiency. We utilize GPT-Neo-2.7B as our foundational model for scoring and inference because it is a distilled model with moderate size. It has good generalization ability and a relatively fast inference speed.
Numerous researchers employing GPT-Neo-2.7B as the base model have consistently demonstrated its effectiveness. \cite{rubin2021learning,li2023unified}

\textbf{Code Llama-13B}
Code Llama-13B \cite{roziere2023code} is an open-source large language model developed by MetaAI, specifically designed for code generation tasks. It has been widely used in previous research. \cite{liu2023agentbench,azerbayev2023llemma}. 
With 13 billion parameters, it leverages advanced natural language processing capabilities to generate code based on given text prompts. Code Llama-13B is part of the Llama series of models and aims to provide developers with a powerful tool for various programming tasks, such as code completion, synthesis, and summarization. Its open-source nature allows for extensive customization and adaptation to different coding environments and requirements.

\subsection{Baselines}
In this paper, we will compare our approach with three baseline methods. 

\textbf{Random}
Random represents the random selection of examples, which is typically used as a vanilla baseline model in ICL work \cite{gao2023makes,nashid2023retrieval} to demonstrate the effectiveness of proposed approaches.

\textbf{BM25}
Best Matching 25 (BM25) \cite{robertson1994some} is a ranking function used in search engines to evaluate the relevance of documents to a search query. It enhances the classic TF-IDF model by incorporating term frequency saturation and document length normalization. BM25 scores are calculated using a formula that balances term frequency with document length, ensuring that frequently occurring terms and varying document lengths are appropriately weighted. It is widely employed in code-related tasks, due to its effectiveness, simplicity, and robustness. \cite{gao2023makes,nashid2023retrieval,ahmed2024automatic} Gao et al. shows that BM25 outperforms pre-trained dense retrievers like UniXcoder \cite{guo2022unixcoder} and CoCoSoDa \cite{shi2023cocosoda}, and achieves state-of-the-art performance in code intelligence tasks.

\textbf{ASAP}
Automatic Semantic Prompt Augmentation (ASAP) \cite{ahmed2024automatic} enhances prompts for code summarization by incorporating three types of designed semantic facts: repository name and path, tagged identifiers, and data flow graphs (DFGs). 
The ASAP pipeline involves configuring a LLM, a pool of exemplars, and a static analysis tool. The process starts by using BM25 to retrieve relevant exemplars, which are then analyzed to extract designed semantic facts. These facts are included in the prompts to LLM, which generates the final output. This method significantly improves performance on the code summarization task by providing LLM with detailed and relevant context.

\begin{table*}[t]

\centering
\caption{Performance of ICR on Code Summarization.}
\label{tab: performance of ICR on cs}
\vspace{-0.2cm}
\begin{tabular}{@{}ccccccccc@{}}
\toprule
                  Tasks  & \multicolumn{8}{c}{Code Summarization}  \\ \midrule
                  Datasets  & \multicolumn{4}{c}{CSN-Java} & \multicolumn{4}{c}{CSN-Python}  \\ \hline

Metrics               & ROUGE-L       &    BLEU-4   &   chrF&    METEOR&   ROUGE-L      &  BLEU-4  &   chrF&METEOR\\  \hline                 
Random + GPT-Neo-2.7B   &   18.5        &      3.6         &    16.8&    17.1& 18.8        &  2.8            & 15.6& 16.0\\
BM25 + GPT-Neo-2.7B     &   20.4       &       4.7        &    19.0&     18.3& 19.5        &  3.2           & 16.5& 16.3\\
ASAP + GPT-Neo-2.7B     &   25.9      &       7.4         &    23.0&      22.5& 20.4       & 4.0          & 17.7& 16.6\\
ICR + GPT-Neo-2.7B      &   \textbf{34.8}       &       \textbf{11.1}        &    \textbf{29.9}&   \textbf{31.8}& \textbf{32.8}       & \textbf{7.6}             & \textbf{26.4}& \textbf{28.4}\\ \hline
Random + Code Llama-13B &  34.0        &       8.1        &  27.7&      28.3& 32.7         &   6.4          & 25.8& 27.1\\
BM25 + Code Llama-13B   &   33.8       &      10.5         & 29.4&       29.5& 34.0          & 8.2         & 28.4& 28.3\\
ASAP + Code Llama-13B   &  35.4        &   11.9            &   29.1&       30.7& 33.0        & 9.3         & 27.1& 27.9\\
ICR + Code Llama-13B    &  \textbf{40.3}        &       \textbf{13.8}        &   \textbf{33.0}&       \textbf{36.1}& \textbf{39.3}        & \textbf{11.0}         & \textbf{30.9}& \textbf{34.1}\\ \bottomrule
\end{tabular}
\vspace{-0.2cm}
\end{table*}

\begin{table*}[h]

\centering
\caption{Performance of ICR on Program Synthesis and bug fixing.}
\label{tab: performance of ICR on pb}
\vspace{-0.2cm}
\begin{tabular}{@{}cccccccc@{}}
\toprule
                  Tasks  &  \multicolumn{3}{c}{Program Synthesis} & \multicolumn{4}{c}{Bug Fixing} \\ \midrule
                  Datasets  &  \multicolumn{3}{c}{Conala} & \multicolumn{2}{c}{B2F\_small} & \multicolumn{2}{c}{B2F\_medium} \\ \hline

Metrics               & CodeBLEU       &    BLEU-4    &   CrystalBLEU      &  BLEU-4       & CrystalBLEU      & BLEU-4     & CrystalBLEU         \\  \hline                 
Random + GPT-Neo-2.7B           &      16.6          &    15.4          &    11.5& 75.0       &  41.7& 87.1        & 48.7\\
BM25 + GPT-Neo-2.7B            &       18.5          &    17.6          &     12.8& 75.6       &  41.9& 86.8        & 48.7\\
ICR + GPT-Neo-2.7B             &       \textbf{32.3}&    \textbf{31.6}&   \textbf{21.4}& \textbf{79.2}       & \textbf{42.1}& \textbf{89.0}        & \textbf{49.1}\\ \hline
Random + Code Llama-13B         &       36.1          &  35.4         &      25.1& 78.2       &   42.0& 88.7        & 48.7\\
BM25 + Code Llama-13B          &      37.1          & 38.0           &       26.4& 78.8       & 42.6& 89.8        & 49.3\\
ICR + Code Llama-13B           &       \textbf{44.4}          &   \textbf{45.8}           &       \textbf{30.4}& \textbf{80.1}       & \textbf{43.5}& \textbf{90.7}        & \textbf{49.7}\\ \bottomrule
\end{tabular}
\vspace{-0.2cm}
\end{table*}

\subsection{Implementation Details}\label{detail}
We trained ICR on $B2F_{small}$, $B2F_{medium}$, CSN-Java, CSN-Python, and Conala datasets.
As with all ICL methodologies, we abstain from updating the large model, focusing solely on retriever training. Due to computational resource constraints, we trained ICR using only a quarter of the CSN-Python dataset, and we trained CSN-Java in four slices to accelerate the training process. In the process of selecting relevant examples for each sample, we opt for K=50 relevant examples per sample. Due to computational limitations, we conducted three rounds of iterative training, and we observed that the loss stopped decreasing after four epochs, so we trained ICR for four epochs in each iteration. Training takes place on a server equipped with 8 NVIDIA A800-80G GPUs. To address the repetition problem, we implemented post-processing to trim the repetitive parts, retaining only distinct instances. Moreover, because we need to train on multiple datasets with significant size differences, the sampling rate for Conala is set to 3 per epoch, while for the other datasets, it is set to 0.7. $|C_m|$ is set to 2048, which is the maximum input length for GPT-Neo-2.7B. Considering that the in-batch-tree loss calculates examples within a batch of queries and ranking-tree loss only calculates examples of the query itself, the in-batch-tree loss is generally larger than the ranking-tree loss. We set $\gamma_1$ to 1 and $\gamma_2$ to 4 to ensure both loss functions are effective. 
In our implementation, we fixed the random seed for consistent results, while employing a greedy decoding strategy when invoking the LLM.

\section{Result Analysis}

\subsection{Effectiveness of ICR}

In RQ1, we compare ICR with BM25 on the three tasks.
Because ASAP only reports its performance on code summarization, we only compare ICR with ASAP in the two code summarization datasets.
For our base model selection, we first compare the performance of ICR with baseline models on the open-source GPT-Neo-2.7B \cite{black2021gpt}.
Table \ref{tab: performance of ICR on cs} and \ref{tab: performance of ICR on pb} presents the evaluation results.
We can observe that on the code summarization task, ICR model significantly outperforms BM25, achieving BLEU-4 score increases of 136.2\% on the CSE-Java dataset and 137.5\% on the CSN-Python dataset. Additionally, it improves over BM25 in ROUGE-L by of 70.6\% and 68.2\% on these datasets, respectively.
Compared to ASAP, a method specifically designed for code summarization, ICR demonstrates improvements of 50.0\% on the CSN-Java dataset and 90\% on the CSN-Python dataset in terms of BLEU-4.
Additionally, ICR achieves increases in ROUGE-L scores of 34.3\% and 60.8\% on these respective datasets. ICR also improves chrF score of over 30\% and METEOR over 41.3\%.

For the program synthesis task, ICR outperforms the BM25 method by 74.6\% on CodeBLEU and 67.2\% on CrystalBLEU. 
Conala is a dataset with a relatively small training set, and when the training set is insufficient, ICR can learn about the quality of examples from other datasets of different tasks. 
In contrast, the small dataset limits the comprehensiveness of the corpus that BM25 can retrieve.
In the Bug Fixing task, ICR outperforms the BM25 method by 3.6 BLEU-4 score on $B2F_{small}$ and 3.2 BLEU-4 score on $B2F_{medium}$, respectively.

To evaluate the generality of our retriever, we transferred ICR trained with feedback from GPT-Neo-2.7B to Code Llama-13B. As shown in Table \ref{tab: performance of ICR on cs} and \ref{tab: performance of ICR on pb}, our model significantly outperforms the baseline methods. 
For the code summarization task, ICR outperforms the BM25 method by 32.4\% on the CSN-Java dataset and 34.1\% on the CSN-Python dataset in terms of BLEU-4, while improving ROUGE-L by 19.2\% and 15.6\%. 
Further, it also surpasses the heuristic-based ASAP method by 16.8\% and 18.3\% on BLEU-4, while improving ROUGE-L by 13.8\% and 19.1\%. Additionally, it achieved the highest scores on chrF and METEOR.
In the program synthesis task, we exceed the BM25 method by 19.7\%. 
In the bug fixing task, ICR's performance also surpasses that of BM25 in BLEU-4 scores. This demonstrates the effectiveness of our ICR when transferred to a more powerful LLM, such as Code Llama-13B.

\find{
When tested with GPT-Neo-2.7B and Code Llama-13B, our method outperforms BM25 and ASAP in all three code intelligence tasks.
}

\begin{table*}[h]
\centering
\caption{Ablation Studies of ICR.}
\label{tab: ablations of ICR}
\vspace{-0.2cm}
\begin{tabular}{@{}ccccccccc@{}}
\toprule
                  Tasks  & \multicolumn{4}{c}{Code Summarization} & \multicolumn{2}{c}{Program Synthesis} & \multicolumn{2}{c}{Bug Fixing} \\ \midrule
                  Datasets  & \multicolumn{2}{c}{CSN-Java} & \multicolumn{2}{c}{CSN-Python} & \multicolumn{2}{c}{Conala} & B2F\_small & B2F\_medium \\ \hline
Metrics               & ROUGE-L       &    BLEU-4    &   ROUGE-L      &  BLEU-4       & CodeBELU      & BLEU-4     & BLEU-4     & BLEU-4      \\  \hline 
ICR + GPT-Neo-2.7B      &   \textbf{34.8}       &       \textbf{11.1}        &    \textbf{32.8}       &   \textbf{7.6}             & \textbf{32.3}& \textbf{31.6}& \textbf{79.2}       & \textbf{89.0}        \\ 
 - w/o Structural Info. &    29.9       &     9.1          &   30.0         &      6.6          & 23.3          &  22.6          & 77.9       & \textbf{89.0}        \\
 - w/o Feedback-driven Training &   30.2        &      7.7         &   21.7         &      3.0          & 32.1          &  28.7          & 65.8       & 88.6        \\  

\bottomrule
\end{tabular}
\vspace{-0.3cm}
\end{table*}

\subsection{Ablation Study}

To evaluate the effectiveness of each component of ICR, we conducted ablation studies on the structural information and iterative training based on LLM feedback. The experimental results are shown in the Table \ref{tab: ablations of ICR}. We have removed two key modules from ICR and constructed the following two variants:

\begin{itemize}[leftmargin=0.3cm]
    \item \textbf{ICR + GPT-Neo-2.7B - w/o Structural Info.}: The ablation of structural information involves whether the structural information is used in calculating similarity and loss. Specifically, we remove the tree structure information from our model and replace it all with zero vectors. 
    \item \textbf{ICR + GPT-Neo-2.7B - w/o Feedback-driven Training}: The feedback-driven training ablation experiment compares single-iteration training and iterative training based on LLM feedback. When building variant of single-iteration, ICR only using BM25 to initialize the examples in the training phase. Therefore, this variant is unable to iteratively train based on the example scores provided by LLM feedback.
\end{itemize}

\subsubsection{\textbf{Structural Information}}

Compared to the variant that does not incorporate structural information, ICR demonstrated an improvement of 22.0\% on CSN-Java and 15.2\% on CSN-Python in terms of BLEU-4 for the code summarization task.
Additionally, CodeBLEU improves 38.6\% on the Conala dataset for the program synthesis task. In the bug fixing task, structural information improves 1.3 BLEU-4 points on $B2F_{small}$. 
Syntax trees contain rich structural information \cite{bai2021syntax}, which can provide valuable cues for code intelligence tasks. 
In addition to characterizing structural information during model input, our method also calculates structural information during training.
The loss function we designed includes structural information that can effectively help the model understand the input structure and retrieve structurally more similar samples.

\subsubsection{\textbf{Feedback-Driven Iterative Training}}

Feedback-driven iterative training has led to significant performance improvements, particularly on the CSN-Java (26.0\% BLEU-4), CSN-Python (153.3\% BLEU-4), and $B2F_{small}$ (20.4\% BLEU-4) datasets. 
For \textbf{ICR + GPT-Neo-2.7B - w/o Feedback-driven Training} variant, given a training sample, it relies solely on the retrieval results from BM25 to select relevant examples for training. This approach could be insufficient and suboptimal.
For ICR + GPT-Neo-2.7B method, we utilize the ICR model trained in the previous iteration for further retrieval. 
Some of retrieved examples might still not be helpful for model inference. This distinction is made through feedback from the large model, training the retriever to discard these unhelpful examples in the next round.
We refine the model based on the feedback from LLM to enhance its performance.
This process represents a effectively continual iteration of corrections and optimizations to ICR.
As the iterations progressed, the performance of ICR improved, enabling the acquisition of higher-quality samples and leading to better overall performance.

\find{
For all three code intelligence tasks, both structural information and feedback-driven training helps ICR find higher quality examples.
}

\begin{table*}[]

\centering
\caption{Performance of ICR on Untrained Datasets.}
\label{tab: untrained of ICR}
\vspace{-0.2cm}
\begin{tabular}{@{}ccccccc@{}}
\toprule
     & \multicolumn{2}{c}{CSN-JavaScript} & \multicolumn{2}{c}{CSN-PHP}  & \multicolumn{2}{c}{CSN-Ruby} \\ \midrule
Metrics & ROUGE-L &  BLEU-4 & ROUGE-L&  BLEU-4   & ROUGE-L &  BLEU-4  \\ \hline
BM25 + GPT-Neo-2.7B & 18.5 &  6.8     & 31.9 & 10.2  & 18.3  &   2.2\\
ASAP + GPT-Neo-2.7B &   22.8 & 9.0   &36.1  &10.7 &  25.1 &   3.9 \\
ICR + GPT-Neo-2.7B  & 23.8 &  10.1   &37.6    & 12.5 & 27.7 & 4.9 \\ \bottomrule
\end{tabular}
\vspace{-0.3cm}
\end{table*}

\begin{table*}[]

\centering
\caption{Analysis of Example Order.}
\label{tab: examples orders}
\vspace{-0.2cm}
\begin{tabular}{@{}cccccc@{}}
\toprule
                   &  $B2F_{small}$ & $B2F_{medium}$  & Conala & CSN-Java & CSN-Python \\ \midrule
Metrics             &   BLEU-4       &      BLEU-4      &   CodeBLEU     &    BLEU-4        &     BLEU-4        \\   \hline                
ICR Random + GPT-Neo-2.7B             &   78.9       &    88.6        &   \textbf{32.6}&      11.1      &         7.5    \\
ICR Similarity + GPT-Neo-2.7B         & \textbf{79.2}       & \textbf{89.0}        & 32.3   &  11.1      &  \textbf{7.6}       \\
ICR Reverse Similarity + GPT-Neo-2.7B & 78.8         &  88.3          &  32.4      &     \textbf{11.2}       &        7.3     \\ \bottomrule
\end{tabular}
\vspace{-0.3cm}
\end{table*}

\subsection{Transferability of ICR on Untrained Datasets}~\label{sec:rq3}
In RQ3, we explore whether ICR can generalize to languages it has not been trained on. 
Considering the CSN dataset includes data from other programming languages, we conducted the experiments on the code summarization task.
We will compare ICR-enhanced GPT-Neo-2.7B with two baselines: BM25 and ASAP.

To answer RQ3, we evaluate the performance of the ICR approach in comparison with baseline models for retrieving examples from untrained programming languages, specifically on the CSN-JavaScript, CSN-PHP, and CSN-Ruby datasets.
We employ ICR that was trained on the CSN-Java, CSN-Python, Conala, $B2F_{small}$, and $B2F_{medium}$ datasets to retrieve similar samples from these untrained datasets. 
For inference, we use GPT-Neo-2.7B, maintaining all settings consistent with those outlined in Section \ref{inference} and Section~\ref{detail}.

While ICR was only trained on Java and Python datasets during training, the results in Table \ref{tab: untrained of ICR} indicate a performance improvement of 48.5\%, 22.5\%, and 122.7\% with ICR compared to BM25 on CSN-JavaScript, CSN-PHP, and CSN-Ruby in terms of BLEU-4, respectively. 
Compared to ASAP, our approach showed improvements of 12.2\%, 16.8\%, and 25.6\% in terms of BLEU-4, respectively.
This suggests that the knowledge acquired by ICR from training data and structural information can be generalized to untrained languages, yielding better performance. 
This is because ICR effectively learns the structural knowledge of code, which is transferable across languages. Additionally, it extracts deeper insights from the feedback of the large language model. 
Unlike BM25, which relies solely on text-based example selection, ICR could provide more relevant and insightful examples. This ultimately offers better inspiration for the large language model.

\find{  
Our approach also shows superior performance on datasets it has not been trained on.
}

\section{Discussion}

\subsection{Findings and Limitations}
Gao et al. \cite{gao2023makes} find that the example selection, quantity, and order are pivotal factors influencing the performance of ICL. While the impact of example selection and example quantity on model performance is widely acknowledged \cite{liu2021makes,zhang2022active}, the effects of order need further discussion. 
Gao et al. observed that while the performance of large models can be affected by the order of examples, this impact diminishes when the quality of examples is high, such as those retrieved using BM25.
For instance, in their study focusing on code summarization enhanced by BM25, randomizing the order of examples resulted in optimal performance.
In general, high-quality examples show less sensitivity to their order~\cite{lu2021fantastically,li2023finding}, this phenomenon that has been confirmed across a range of natural language processing tasks. 

To evaluate the effect of example order, we compare random order and two simple ordering methods, i.e., ICR similarity order and ICR reverse similarity order. 
The ICR similarity method organizes examples based on their retrieval order, as described in Section \ref{inference}, placing those with higher similarity closer to the input query. 
Conversely, the ICR reverse similarity order arranges examples so that those with lower similarity are positioned nearer to the input query.
As illustrated in Table \ref{tab: examples orders}, reversing the order of ICR examples fed into the model has no substantial impact on the results for most datasets. This finding indirectly indicates the high quality of the examples retrieved by ICR and consistently provides valuable insights into the model, making the order of examples relatively unimportant.

We posit that the fundamental factors impacting the augmentation of large models through ICL lie in example selection and composition. While each individually chosen example may be optimal, their collective efficacy may be compromised due to potential redundancy. Hence, opting for a more varied selection of examples may enhance the inference capabilities of large models. However, an exhaustive examination of all possible example combinations is computationally prohibitive, necessitating the use of approximate algorithms in this regard. Our current study does not delve into example composition, leaving this avenue for future investigation.

\subsection{Threats to validity}

\noindent\textbf{Threats to internal validity}
To enable ICR to perform multiple tasks, we employed different instructions for each task. The variation in instruction selection may impact model performance. To mitigate this threat, we tested the performance of different instructions on a small-scale subset of the corresponding training set and select the best-performing instruction.

\noindent\textbf{Threats to external validity}
ICR adopts a unified architecture for multitask learning, with training conducted solely on three code-related tasks and datasets in two languages, i.e. Python and Java. Performance on the datasets in other languages and in other tasks remains to be tested in future evaluations. To mitigate this threat, we assessed the performance of ICR on datasets in other languages that were not part of its training data (see Section~\ref{sec:rq3}). Another threat is that Conala is a relatively small data set. This may lead to insufficient training in the program synthesis task. Therefore, it may not be able to retrieve enough high-quality examples. Additionally, this threat could potentially compromise the evaluation of ICR on the code synthesis task, making it less generalizable.
In the future, we plan to conduct experiments on more datasets for program synthesis task.

\noindent\textbf{Threats to construct validity}
For the same dataset, we utilized the evaluation metrics identical to those used by Gao et al. \cite{gao2023makes}, and the metric we used is a potential threat, as it may not fully reflect the quality of the output. We did not conduct manual evaluations of ICR, which we plan to undertake in the future.

\section{Related Works}

\subsection{Pre-Trained Models for code intelligence tasks}

Pre-trained models have achieved significant accomplishments in the field of software engineering. Drawing inspiration from the triumph of large pre-trained models in NLP~\cite{brown2020language,devlin2018bert,liu2019roberta,touvron2023llama}, Feng et al. \cite{feng2020codebert} adapted a pre-trained language model named CodeBERT, leveraging a vast code corpus. Guo et al. \cite{guo2020graphcodebert} further enhanced this approach with GraphCodeBERT, integrating structural code information for superior performance. Both CodeBERT and GraphCodeBERT have excelled in programming language tasks. Ahmad et al. \cite{ahmad2021unified} mirrored BART's \cite{lewis2019bart} model structure, pre-training their PLBART model, tailored for bug fixing. Meanwhile, Wang et al. \cite{wang2021codet5} introduced CodeT5, a pre-trained code model fashioned after T5's \cite{raffel2020exploring} architecture, along with specialized pre-training tasks for code. CodeT5 has demonstrated outstanding performance across various downstream tasks, including code generation. These seminal models represent significant strides towards leveraging pre-trained models for code-related tasks, demonstrating remarkable efficacy across diverse downstream applications.

While pre-trained models offer a myriad of benefits, they also come with their fair share of limitations that can hamper their real-world utility. One significant drawback is the mismatch between the pre-training tasks and downstream applications, requiring fine-tuning to bridge the gap. This fine-tuning process can be cumbersome for users, adding complexity to model deployment. Moreover, due to their relatively limited parameter space, pre-trained models often struggle with generalizing across diverse tasks, resulting in a lackluster performance on tasks and datasets that haven't undergone specific fine-tuning procedures.

\subsection{Large Language Models}

With the further development of pre-trained models, models with larger parameters have emerged, known as Large Language Models (LLMs). LLMs demonstrate exceptional generalization capabilities across a wide range of tasks, eliminating the need for fine-tuning to achieve good performance. The GPT series of models developed by OpenAI, including GPT-3 \cite{brown2020language}, ChatGPT, GPT-4 \cite{achiam2023gpt}, CodeX \cite{chen2021evaluating}, and others, have achieved success in various tasks. GPT-Neo \cite{black2021gpt} is an open-source model distilled by EleutherAI from GPT-3, possessing fewer model parameters yet achieving comparable performance to larger models. 
Code Llama \cite{roziere2023code} is an open-source code generation model released by MetaAI, capable of generating code using text prompts.

The advent of large language models has liberated users from the complexities of fine-tuning models. However, these models are particularly sensitive to their input prompts \cite{zhu2023promptbench}. Hence, the current challenge revolves around creating prompts of superior quality for these large language models. In-Context Learning (ICL) technology steps in to tackle this challenge by furnishing examples to inspire and guide these expansive models.

\subsection{In-Context Learning}

Large Language Models (LLMs) have demonstrated remarkable emergent capabilities, and enhancing them has become a focal point in research. In-context learning (ICL) has shown a series of successes in augmenting large language models. In ICL, the model completes new tasks directly by being provided with a set of contextual examples, without the need for further fine-tuning or parameter modification. These contextual examples typically include input-output pairs, which the model uses to infer the output for new inputs. Brown et al. \cite{brown2020language} were among the pioneers to showcase that providing a series of examples can significantly boost the performance of GPT-3. 
DPR \cite{karpukhin2020dense}, employing a dual-encoder structure and a contrastive learning approach, has achieved notable success in open-domain question-answering tasks. 
Gao et al. \cite{gao2023makes} were the first to apply ICL techniques to code-related tasks, utilizing BM25 \cite{robertson1994some} to retrieve examples and enhance large language models in code intelligence tasks. Ahmed et al. \cite{ahmed2024automatic} further utilized semantic facts to augment the performance of the code summarization task after BM25 retrieving examples.

However, in the field of code intelligence tasks, existing approaches are still based on lexical similarity retrievers~\cite{gao2023makes, ahmed2024automatic}.
In this paper, we proposed a novel approach ICR based on contrastive learning to help retrieve highly relevant examples for a given code task.
Our approach incorporates both semantic and structural information to help retriever understand the relationship between queries and examples.
This retriever will inspire LLMs to generate more accurate results to complete various code intelligence tasks.

\section{Conclusion and Future Work}

In this paper, we introduced an approach named Instructive Code Retriever (ICR), which is designed to fetch high-quality examples for multiple code intelligence tasks, thereby enhancing the inference of large language models. Given the training data, ICR first uses BM25 (iter 0) or the iteration-trained ICR to score examples and search for relevant ones. After getting the relevant examples, we further leverage LLMs to rank them based on their generation probability.
Then we use this ranking to calculate a customized tree-based loss function to achieve the goal of training ICR. During testing, ICR effectively retrieved useful examples to inspire the large model. Across various code intelligence tasks, such as bug fixing, program synthesis, and code summarization, ICR demonstrated significant improvements over baseline models. Additionally, we analyzed each component of ICR and assessed its transferability across datasets in different languages. Overall, our retriever shows promise in assisting researchers and developers across diverse code intelligence tasks and datasets in various languages. In the future, we aim to delve deeper into the impact of example combinations and diversity on inspiring large models, as well as enhancing retrievers' ability to learn code-specific structures. Our replication package is available at https://github.com/kingofheven/ICR.

\begin{acks}
This research is supported by the National Natural Science Foundation of China (No. 62202420 and No. 62302430), the Fundamental Research Funds for the Central Universities (No. 226-2022-00064), and Zhejiang Provincial Natural Science Foundation of China (No. LQ24F020017). Haoye Wang is supported by Zhejiang Provincial Engineering Research Center for Real-Time SmartTech in Urban Security Governance.
\end{acks}

\balance
\bibliographystyle{ACM-Reference-Format}
\bibliography{sample/ASE/mybib}

\end{document}